# A Comparative Study of Data Storage and Processing Architectures for the Smart Grid

María Arenas-Martínez
Sergio Herrero-Lopez
Abel Sanchez
John R. Williams
Massachusetts Institute of Technology
Cambridge, Massachusetts 02139
{marenas, sherrero, doval, jrw@mit.edu}

Paul Roth
Paul Hofmann
SAP Labs
Palo Alto, California 94304
{paul.roth, paul.hofmann}@sap.com

Alexander Zeier
Hasso Plattner Institute for
Software Systems Engineering
Potsdam, Germany
{zeier}@hpi.uni-potsdam.de

*Abstract*—A number of governments and organizations around the world agree that the first step to address national and international problems such as energy independence, global warming or emergency resilience, is the redesign of electricity networks, known as *Smart Grids*. Typically, power grids have "broadcasted" power from generation plants to large population of consumers on a suboptimal way. Nevertheless, the fusion of energy delivery networks and digital information networks, along with the introduction of intelligent monitoring systems (*Smart Meters*) and renewable energies, would enable two-way electricity trading relationships between electricity suppliers and electricity consumers. The availability of real-time information on electricity demand and pricing, would enable suppliers optimizing their delivery systems, while consumers would have the means to minimize their bill by turning on appliances at off-peak hours. The construction of the *Smart Grid* entails the design and deployment of information networks and systems of unprecedented requirements on storage, real-time event processing and availability. In this paper, a series of system architectures to store and process *Smart Meter* reading data are explored and compared aiming to establish a solid foundation in which future intelligent systems could be supported.

## I. INTRODUCTION

Currently, the design and size of an electricity supply network corresponds to the demand observed during peak periods. Unfortunately, this leads to an overdimensioned network for the rest of the off-peak periods. Therefore, one of the *Smart Grid*'s goals is to modify customer behavior patterns in order to balance the load and reduce peak electricity demand. A naïve way to motivate this behavior shift is to apply tariffs based on the time of usage, setting different electricity prices for peak and off-peak periods. Even the simplest pricing mechanism requires information about the time and volume of electricity consumed by every customer. For this purpose, *Smart Meters* (SMs) will be installed in every household participating in the *Smart Grid*. These devices will periodically measure and temporarily store each customer's consumption data, then this information will be transferred to the utility companies through the Advanced Metering Infrastructure (AMI) [1].

Furthermore, the progressive integration of in-plug hybrid vehicles (PHEVs) and in-plug electric vehicles (PEVs) can dramatically change the way customers consume electricity creating sharper consumption peaks and usage patterns never seen before. The interaction between electric vehicles, renewable energies and urban areas has been recently explored. The possibility of electric vehicles trading and storing electricity with private buildings and stores has been promoted by researcher communities [2]. In any case, envisioned use cases strengthen the argument that a scalable, reliable and fast information infrastructure is required in order for the *Smart Grid* to accomplish its objectives.

This paper presents a series of system architectures that aim to provide a solid foundation for the data consumption systems that will make the *Smart Grid* an intelligent network. The organization of this paper is as follows. Section II introduces related work on this research area. Section III and IV are the core of this paper: Section III presents different system architectures to store and efficiently process the new volume of data from the SMs, and Section IV explains the basics of the simulator we have built to test the proposed solutions in Section III. Section V summarizes the simulation result for each system architecture and finally, Section VI presents the conclusions of this work.

## II. RELATED WORK

Since the Obama administration stimulated advanced electricity grid projects by unveiling a 3.4 billion grant, a series of initiatives have been proposed to create the new *Smart Grid*. In this section, some of these are introduced:

The *Mobility-On-Demand* (MoD) [2] project proposes to integrate electric vehicles into the design and construction of modern cities. This approach considers the electric vehicle to be the keystone element of the future electric grids and the need to establish a symbiosis between the vehicles and the surrounding buildings which will soon became electricity suppliers due to the installation of renewable energy generators. MoD researchers envision an ecosystem in which vehicles and appliances will trade energy with sources that will be an integral part of urban areas.

The *Tennessee Valley Authority* (TVA) [3] has constructed a *Smart Grid* information processing architecture using the Map-

Reduce paradigm [4] and the Hadoop Open Source project [5]. They are capable of processing hundreds of TB of data from power grids, in order to detect power grid anomalies, creating power grid maps and evaluating power consumption history. Unfortunately, by design, Hadoop is a batch processing system and it is not meant to execute high-performance low latency queries and would not work for real-time applications.

Since March 2008, *Xcel Energy* Utility company has been collaborating in the construction of a *Smart Grid* pilot infrastructure in Boulder, Colorado [6]. With 100.000 habitants and being the home to the *University of Colorado* and federal institutions like the *National Institute of Standards and Technology* (NIST), Boulder is set up to be the world's first *Smart Grid* city. *Xcel Energy* is also pioneer in launching dynamic pricing pilots, offering participants different tariffs as an incentive to shift electricity consumption from on-peak to off-peak periods. Three pricing options are being tested: Time-Of-Use (TOU) rate, Critical Peak Pricing (CPP) rate and Peak Time Rebate (PTR) rate. The TOU rate will break the day into two periods (on-peak and off-peak); CPP Rate will add a third interval to the TOU rate when system capacity/economic conditions require reduced energy usage; and PTR rate which allows customers paying the standard residential rate and at the same time offers a rebate if the electricity consumption is reduced during critical peak times.

None of these efforts faces the challenge of storage nor computation, therefore we consider that our contribution is ortogonal to these projects and collaborates to strengthen the vision of the *Smart Grid*.

## III. Storage and Processing Architectures

### A. Introduction

Different architecture designs are described in this section. The components of these architectures are summarized below.

- *Concentrator* node (CN): A Concentrator node gathers, stores and returns electricity consumption data from multiple SMs. It is a passive node in the sense that receives and executes orders. At the end of the day, the CN is asked to collect the SM readings from each household linked to it and stores the data so that it can be processed later on demand. The communication protocols between the Concentrator Node and the SMs are specified by AMI [1].
- *Central Data Processing* node (CDPN): The Central Data Processing node manages the CNs. Being the only active node in the system, it constitutes the highest level of control. It is responsible for managing and coordinating the tasks assigned to CN as well as calculating electricity consumption statistics and monthly billing. Besides, the CDPN generates a set of *Time Buckets* to calculate the monthly bill. A *Time Bucket* is defined as: "a continuous or intermittent period of time in which all the SM readings have the same price". An example of a *Time Bucket* is the following: "Saturdays and Sundays, from 0:00 AM to 5:59 AM between the January 1, 2009 and January 31, 2009". This *Time Buckets* are required for TOU and CPP pricing types.

### B. Architecture I: Single Relational Database

This architecture is composed by one CDPN, a set of CNs and one Relational Database Management System (RDBMS) located at the CDPN.

The database schema consists of a single table with all the SM readings collected by each CN. Each row uniquely identifies each SM reading and stores the electricity consumption in kWh for a specific time interval. Every day, the CDPN sends a request, called *Collect*, for collecting the electricity consumption data from each CN. Each CN receives via AMI the daily collection of SM readings per household, keeps the data in a memory buffer and then *Inserts* the SM readings into the database using one single transaction. These batch inserts are queued and executed sequentially against the database. At the end of the month, the total amount of SM readings is stored in the database and the CDPN computes the bill following the next steps: (1) Generates the set of *Time Buckets*; (2) Produces the billing SQL query string; (3) Executes the SQL query against the database. This query classifies each read into its corresponding *Time Bucket* and computes the total price for every household. This is illustrated in Figure 1.

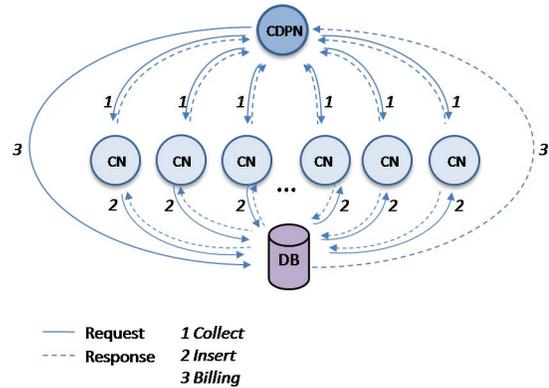

Fig. 1. Single Database Storage

### C. Architecture II: Distributed Relational Database

In this architecture, an RDBMS per CN was considered in order to provide parallel database access and to reduce the excessive database size observed in Architecture I. The design of each RDBMS is identical to the schema described in Architecture I.

Analogous to Architecture I, every CN is asked to collect and store the electricity consumption data at the end of the day. As each CN is associated to a different RDBMS, the time it takes to store the data is now reduced by the number of CNs (*#CN*). Besides, the size of each database is also *#CN* times smaller. On the other hand, since data is distributed across databases and the CDPN does not have direct access to each of them, the CDPN sends the billing query to each CN and then these are responsible for executing the query against the corresponding database and forwarding the results to the CDPN. This architecture is illustrated in Figure 2.

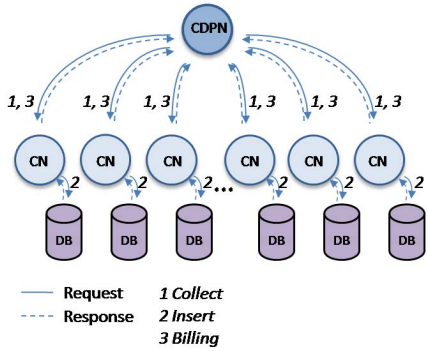

Fig. 2. (Key-Value) Distributed Database Storage

### D. Architecture III: Key-Value Distributed Database Storage

In this architecture instead of writing a database row per SM reading, we store all monthly readings for each household into a single row composed by a household identifier and an xml structured string. Readings and timestamps are appended to this string as they are generated. The storage of xml strings in databases as key-value pairs is commonly utilized [7]. Analogous to Architecture I and II, every day each CN collects the SM readings per household. Data is conveniently stored in the CDPN node memory buffer, while the CNs update the database by appending the xml structured string of each household to its corresponding column.

Due to the increased complexity of the SQL query, the electricity bill computation was no longer executed within the database, but in-memory at the CDPN node. For this purpose, an in-memory *Multi-Core Billing (MCB)* algorithm was designed.

*1) In-memory Multi-Core Billing (MCB) Algorithm:* An inherent advantage of storing the electricity consumption information in-memory at the CDPN node is the possibility of utilizing state-of-the-art multi-core processors to accelerate data processing. In this project, an algorithm that aggregates SM reading data into *Time Buckets* was developed. For each monthly collection of SM readings, each read is classified to the corresponding bucket and then the SM readings within the same *Time Bucket* are aggregated. The algorithm works as follows: The entire dataset is divided by household into $N$ partitions, where $N$ is the number of processing cores available, then each core receives one of these partitions. For each household, the thread assigned to the processing core follows a two step process. 1) *Sort Phase*: The thread sorts the SM readings so that those that belong to the same *Time Bucket* are contiguous in memory. 2) *Aggregate Phase*: The thread takes the sorted values for each bucket and carries out a reduction operation. The reduction of data residing in contiguous locations of the memory is an extremely fast operation due to data prefetching.

In order to accelerate the sorting phase the use of expensive conditional statements (if) is avoided, and an indirection array is introduced instead. The indirection array, called *Mask*, contains the index of the location of each SM in the sorted array. The *Mask* is associated to each *Time Bucket* definition and is pre-calculated once for all the SMs. The algorithm is illustrated in Figure 3.

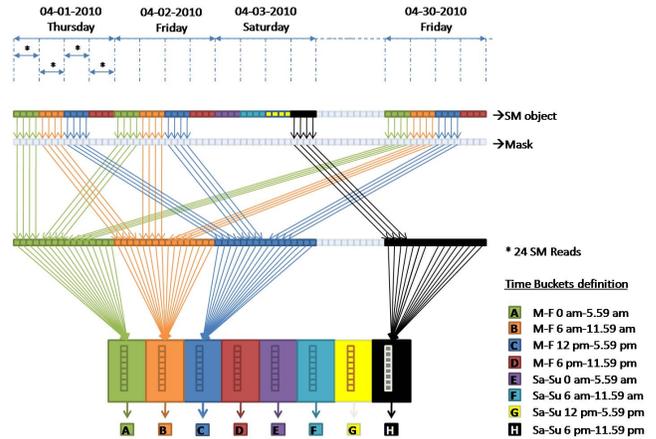

Fig. 3. In-Memory Multi-Core Billing (MCB) Algorithm

### E. Architecture IV: Hybrid Storage (RDBMS and FS)

Since the aggregation query is no longer executed in the database, another storage approach can be proposed based on the combination of a File System (FS) and a RDBMS. As it is illustrated in Figure 4, this architecture is composed by one CDPN with a single database and a set of CNs equipped with their local File Systems.

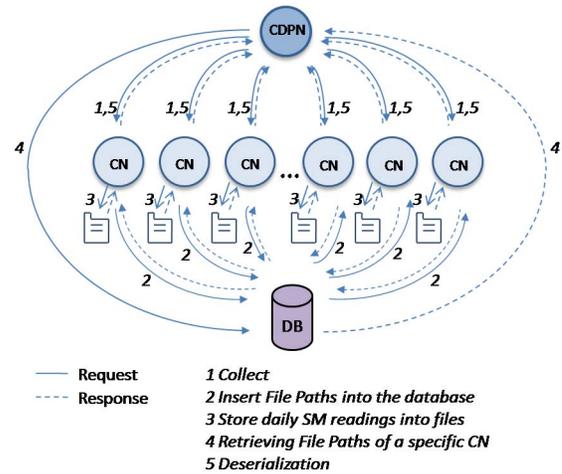

Fig. 4. Hybrid Storage

In this approach the SM reading information is not stored in the database, but in xml files in the FS instead. The database stores pointers to the files in the FS. The database consists of a single table and two columns: one with the pointer to the xml file containing the timestamp information and the other with the pointer to the xml file containing the electricity consumption.

On the other hand, Figure 5 illustrates the structure of the File System design used. As it is shown in the figure, the folders form a hierarchical tree structure composed by 3 levels. The first level groups all the files that are generated in the same month and year while the second and third level are used to distribute the files into organized folders. We used the local File System for convenience, nevertheless a Distributed File System, such as HDFS [8], could also be used for this architecture.

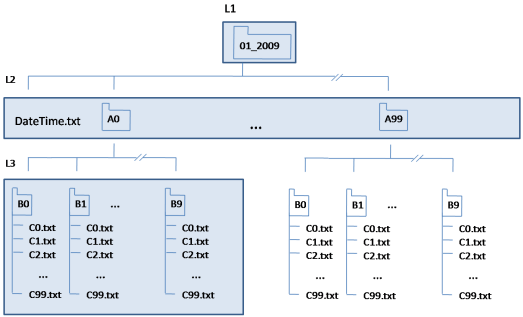

Fig. 5. File System tree

As one would expect, every day CDPN asks each CN to collect the data from each household. The first day of each month, each CN generates a new root folder of Level 1 and builds the described hierarchical tree structure under this folder. Then it creates the same number of files as households plus one file that will be shared by every household and that will contain the date and time information. The relationship between the files paths and the SMs is finally stored in the database. Similarly to Architecture I, *Insert* request are queued and executed sequentially. An advantage of this architecture is that high traffic to the database is not generated as the file path insertion is executed monthly and the number of records per CN is set by the number of households. Once the FS is built and the CN has collected daily data, the CN serializes and stores it in files, while keeping a copy on the CDPN memory buffer. Finally, at the end of the month the CDPN executes the in-memory *MCB* algorithm.

## IV. SIMULATION

### A. Message-Passing Architecture

In order to test the architectures proposed in Section III, we have implemented a *Smart Grid* simulator that reproduces the interactions between CNs, CDPN nodes and databases. The foundations of the simulator take multiple concepts from the Erlang programming language [9].

- Individual *Smart Grid* components follow an Actor model [10] [11], where Actors use message-passing to communicate. Actors with no messages to process block no CPU threads.
- The non-blocking nature of message-passing gives Erlang type models reputation for scalability on multi-core machines. Actors with work to do are scheduled efficiently across available resources.
- Processing one message at a time prevents concurrency problems between Actors and increases stability.

Our simulator was constructed using two components: Concurrency and Coordination Runtime (CCR) and Decentralized Software Services (DSS). 1) Concurrency and Coordination Runtime (CCR) [12], is a set of low-level asynchronous message passing primitives. One of the simplest is *Receive* which is analogous to the Erlang primitive. But there are more sophisticated primitives, such as *Join* (receive a message all of some channels) and *Choice* (receive a message from any of some channels), which can be nested and composed in a number of different ways. These primitives are also non-blocking. *Receivers* produce tasks to handle the messages. Tasks are pushed into queues called *Ports*, which are serviced by a small number of threads. 2) Decentralized Software Services (DSS) [13] is the programming model, which is built on top of the CCR. It is composed by DSS Services, which mimic Erlang processes, and mandates asynchronous message-passing for inter-service communication. Like in Erlang, there is no difference between invoking local services or remote services, although the latter one requires (de)serialization. A detailed description of a DSS Service is provided in section IV-B.

### B. Decentralized Software Services

Every service consists of a *contract*, an internal *state*, *behaviors* and execution *context*. The *contract* identifies the service in a globally unique way using a URI (Universal Resource Identifier) and defines the messages that can be sent to the service; the internal *state* encompasses all the properties (permanent and variable) that the service requires in order to control its own operation; the *behaviors* are the collection of operations that the service can perform to achieve its purpose; and the execution *context* refers to the communication with other services defined as partners, as well as with its initial state.

Each behavior is defined within a *port* and is characterized by a unique request message. A message is a class or data type associated with the port. When a request message is posted to the port, it is delivered to the corresponded behavior and once the algorithm is executed, a response message is sent back to the service. Behaviors are implemented by handlers and are classified as *exclusive* or as *concurrent*. Concurrent handlers can be executed simultaneously with other concurrent handlers, but not while exclusive handlers are running. Figure 6 illustrates a service's components as well as the message-passing between services.

### C. Implementation

The two main building blocks, the CDPN and CN nodes, are both implemented as DSS services in C#: the CDPN-Service and the CN-Service. CDPN-Service is partnered with every CN-Service, in order to send/receive messages to/from each CN-Service as it is illustrated in Figure 6. We chose *SQL Server 2008* to be our RDBMS. In order to establish communication between the CDPN- and CN-services with the database, we created a database front-end service called, *SQL Client Service*. In contrast to other DSS Services, *SQL Client Service* has two message dispatcher queues and two threads, one to deal with database transactions and the other to manage requests from other services. *SQL Client Service* uses the ADO.NET framework [14] to insert bulk load data into the database. For an optimized performance, ADO.NET is configured to perform the operations in a single batch, non-transacted way and minimally logged.

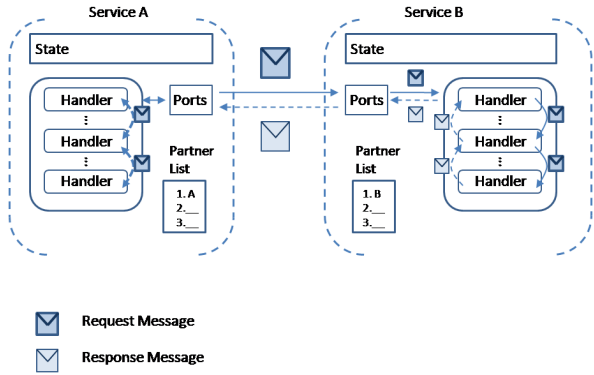

Fig. 6. Service components

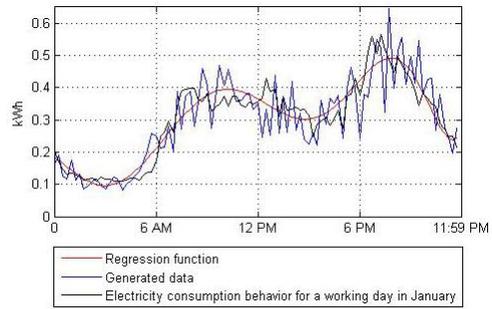

Fig. 7. Generation of one household's daily SM Readings

The simulator runs on a host machine where multiple Virtual Machines (VM) are set up. The CDPN node runs in the host machine and each CN node in a Virtual Machine.

*D. Smart Meter Data Generation*

In order to simulate the collection of data samples from the SMs, tenth grade regression functions where constructed from information collected from a real households. We calculated a set of regression functions that are different depending on the month of the year, and distinguishing workdays from weekends. In order to simulate different households, random noise was added to the values generated by the regression functions.

Figure 7 shows an example of the generated data for one household in a working day in January. The figure compares the generated data with a real working day in January as well as with the regression function obtained for that month.

## V. EXPERIMENTS - RESULTS

This section presents the results obtained during the simulation phase. Every simulation stores and processes SM reading data of an entire month (31 days) for a specified number of households per CN node. The amount of data that a CN node has to handle per household is 2976 SM readings/month (96 SM readings/day) as SM readings are defined to indicate electricity consumption during 15 min intervals. The purpose is to find the architecture which could scale to large number of households per CN node and which would enable processing the data in near real-time. On the other hand, every bill calculation is based on the collection of 8 *Time Buckets* described in Figure 3. It is important to indicate that *Time Bucket* definitions are not available until the month has concluded, therefore, preaggregation is not an option in this study. Table I contains the characteristics of every machine used during this simulation phase. Throughout the section, unless other specifications are mentioned, every CDPN node has been run in Host Machine #1 and every CN node has been run in the VM described in Table I.

TABLE I
HARDWARE SPECIFICATIONS

| Host Machine #1 | Intel(R) Xeon (R) CPU E5345 @2.33 GHz (2 processors, 8 cores, 8 threads), 24 GB RAM Windows Web Server 2008 R2 (64-bit) |
|---|---|
| Host Machine #2 | FourSix Core E7450Xeon @2.4 GHz (24 processors, 24 cores, 24 threads), 64 GB RAM Windows Web Server 2008 R2 (64-bit) |
| Virtual Machine | Intel (R) Xeon (R) CPU E5345 @2.33 GHz (2 processors, 2 threads), 4 GB RAM, 50 GB Hard Disk, Windows VistaTM Enterprise (64-bit) |

Architecture I was simulated for the simple scenario of having one CDPN node and one CN node. Increasing the number of households per CN node followed a linear behavior and it turned out that the VM hosting the CN node could handle up to 100,000 households. Under this scenario, storing daily electricity consumption data into the database took $\sim 1$ min and executing the monthly bill query $\sim 20$ min. Additionally, simulating two CN with 100,000 households each stored daily data of each CN node concurrently in $\sim 1$ min and took $\sim 36$ min to process the monthly bill. However, increasing the number of CN nodes decreased the performance of inserting the data into the database as the CDPN node was not able to insert concurrently CN nodes' data.

Architecture II also followed a linear behavior up to 100,000 households and was simulated for multiple CN nodes with 100,000 households each. It turned out that the maximum number of CN nodes that the host machine could handle were 3. Storing daily electricity consumption data of all the

CN nodes took ∼ 1 min and executing the billing query ∼ 150 min, as if there was only one CN node. Therefore, there was a significant improvement in terms of data storage as this architecture could scale out for multiple CNs nodes if only storage requirements were considered. However, regarding billing computation, the performance decreased drastically as it was carried out by each CN node which was hosted by a VM.

In Architecture III, the maximum number of households that a CN node could manage was 4,000. Appending the xml structured string to the database consumed all the memory resources. In addition to this, the performance decreased as the size of the column was getting larger. Hence, we concluded that this architecture did not scale and was not simulated for multiple CN nodes.

Architecture IV was tested considering one CDPN node and one CN node with 100,000 households. It took ∼ 2 min to write the CN's collected daily electricity data into the File System. Besides, the process of deserializing all monthly data from disk on a "cold start" was measured, and resulted on ∼340 min. This architecture was also tested for 3 CN nodes, where storing the data into the File System took ∼ 2 min for each CN node.

Finally, we explored the multi-core scalability of our *Multi-Core Billing (MCB)* algorithm. In this case Host Machine #2 was used. We carried out experiments for different numbers of households and different numbers of threads. Figure 8 describes the speedup produced by the use of multi-core technology and Table II the timings registered for different problem sizes and threads.

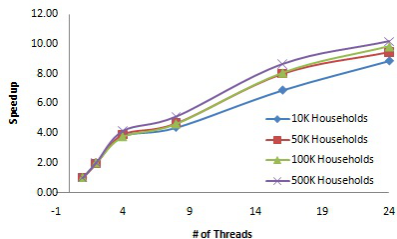

Fig. 8. Multi-Core Billing (MCB) Algorithm's performance analysis

TABLE II
IN-MEMORY MULTI-CORE BILLING (MCB) ALGORITHM'S
PERFORMANCE FOR DIFFERENT # HOUSEHOLDS AND THREADS.

| #Threads | 10K | 50K | 100K | 500K |
|---|---|---|---|---|
| 1 | 3.72 s | 18.25 s | 36.5 s | 190.24 s |
| 2 | 1.86 s | 9.28 s | 18.36 s | 94.33 s |
| 4 | 0.98 s | 4.66 s | 9.71 s | 45.81 s |
| 8 | 0.85 s | 3.91 s | 7.89 s | 37.18 s |
| 16 | 0.54 s | 2.28 s | 4.53 s | 20.01 s |
| 24 | 0.42 s | 1.93 s | 3.71 s | 18.71 s |

From these results, we concluded that the hybrid RDBMS and File System solution resulted on better scalability and performance on the storage of SM reading information. We also showed that the aggregation of SM reading data into *Time Buckets* in-memory and using multi-core technology, outperforms SQL query execution on the database.

## VI. CONCLUSIONS AND FUTURE WORK

Throughout this paper different system architectures have been studied in order to provide a solid foundation for the storage and processing of *Smart Grid* data. We also showed the benefits of processing SM reading data in-memory making the most of state-of-the-art multi-core processors. Unlike batch processing systems, such as Hadoop, in-memory multi-core processing can accelerate the calculation of prices and readjustments of tariffs in order to approach the real-time goal of the *Smart Grid*. We have also shown that storing the data on text files, while keeping the database for metadata on this files, provides horizontal scalability for the system, which cannot be achieved by classic RDBMS.

In the future, we plan to combine the Hadoop architecture with multi-core execution of other dynamic pricing strategies aiming to create a hybrid scalable batch and real-time information systems for the *Smart Grid*.

## ACKNOWLEDGMENT

The authors would like to thank SAP Labs for sponsoring the research. This work was also supported by the EJ/GV Researcher Formation Fellowship BFI.08.80. Also George Chrysanthakopoulos and Henrik Nielsen of Microsoft Research for their assistance with Robotics Studio and the CCR.